\newcommand{\beq}{\begin{equation}}
\newcommand{\eeq}{\end{equation}}
\newcommand{\bea}{\begin{eqnarray}}
\newcommand{\eea}{\end{eqnarray}}

\newcommand{\gsim}{\lower.7ex\hbox{$\;\stackrel{\textstyle>}{\sim}\;$}}
\newcommand{\lsim}{\lower.7ex\hbox{$\;\stackrel{\textstyle<}{\sim}\;$}}

\documentclass[aps,prev,twocolumn,preprintnumbers,floatfix,nofootinbib]{revtex4-1}
\usepackage{graphicx}
\usepackage{epstopdf}
\usepackage{mathrsfs}
\usepackage{amssymb}
\usepackage{verbatim}

\def\stacksymbols #1#2#3#4{\def\theguybelow{#2}
    \def\vp{\lower#3pt}
    \def\sp{\baselineskip0pt\lineskip#4pt}
    \mathrel{\mathpalette\intermediary#1}}

\def\intermediary#1#2{\vp\vbox{\sp
     \everycr={}\tabskip0pt
     \halign{$\mathsurround0pt#1\hfil##\hfil$\crcr#2\crcr
              \theguybelow\crcr}}}

\def\be{\begin{equation}}
\def\ee{\end{equation}}
\def\bea{\begin{eqnarray}}
\def\eea{\end{eqnarray}}

\def\s{\sigma}

\def\sp{\;\;\;,\;\;\;}

\def\lsim{\raise0.3ex\hbox{$\;<$\kern-0.75em\raise-1.1ex\hbox{$\sim\;$}}}
\def\gsim{\raise0.3ex\hbox{$\;>$\kern-0.75em\raise-1.1ex\hbox{$\sim\;$}}}

\def\s{\smallskip}

\def\inbar{\,\vrule height1.5ex width.4pt depth0pt}

\def\IC{\relax\hbox{$\inbar\kern-.3em{\rm C}$}}
\def\IQ{\relax\hbox{$\inbar\kern-.3em{\rm Q}$}}
\def\IR{\relax{\rm I\kern-.18em R}}
 \font\cmss=cmss10 \font\cmsss=cmss10 at 7pt
\def\IZ{\relax\ifmmode\mathchoice
 {\hbox{\cmss Z\kern-.4em Z}}{\hbox{\cmss Z\kern-.4em Z}}
 {\lower.9pt\hbox{\cmsss Z\kern-.4em Z}}
 {\lower1.2pt\hbox{\cmsss Z\kern-.4em Z}}\else{\cmss Z\kern-.4em Z}\fi}

\def\comment#1{}
\def\to{\rightarrow}

\def\u1x{U(1)_X}
\newcommand{\nc}{\newcommand}
\nc{\LL}{L}
\nc{\vv}{\tilde{v}}
\nc{\ccdot}{\!\cdot\!}
\nc{\gsm}{G_{SM}}
\nc{\vfive}{\mathbf{5}\oplus\mathbf{\overline{5}}}
\nc{\vten}{\mathbf{10}\oplus\mathbf{\overline{10}}}
\nc{\zhol}{Z^{\rm hol}}
\nc{\xfb}{\,{\rm fb}}

\begin{document}

%
%

\preprint{DESY 11-240}
\preprint{LPT--Orsay 11/117}

\vspace*{1mm}

\title{Implications of LHC searches for  Higgs-portal dark matter  }

\author{Abdelhak Djouadi$^{a,b}$}
\email{abdelhak.djouadi@th.u-psud.fr}
\author{Oleg Lebedev$^{c}$}
\email{lebedev@mail.desy.de}
\author{Yann Mambrini$^{a}$}
\email{yann.mambrini@th.u-psud.fr}
\author{J\'er\'emie Quevillon$^{a}$}
\email{jeremie.quevillon@th.u-psud.fr}

\vspace{0.1cm}
\affiliation{
${}^a$ Laboratoire de Physique Th\'eorique 
Universit\'e Paris-Sud, F-91405 Orsay, France.
 }
 \affiliation{
${}^b$ CERN, CH--1211, Geneva 23, Switzerland.
}
 
\affiliation{
${}^c$  DESY Theory Group,  Notkestrasse 85, D-22607 Hamburg, Germany.
}

\begin{abstract} 

The search for the a Standard  Model Higgs boson at the LHC is reaching a
critical stage as the possible mass range for the particle has become extremely 
narrow and some signal at a mass of about 125 GeV is starting to emerge.   We
study the implications of these LHC Higgs searches  for Higgs-portal models of
dark matter in a rather model independent way. Their impact on the cosmological 
relic density and on the direct detection rates are studied in the context of 
generic scalar, vector and fermionic thermal dark matter particles. Assuming a
sufficiently small invisible Higgs decay branching ratio,  we find that current 
data, in particular from the XENON experiment, essentially exclude fermionic
dark matter as well as light, i.e. with masses below $\approx 60$ GeV, scalar
and vector dark matter particles. Possible observation of these particles at the
planned upgrade of the  XENON experiment as well as in  collider searches is
discussed.  

\end{abstract}

\maketitle


\section*{Introduction}

The ATLAS and CMS collaborations have recently reported on the search of  the
Standard Model (SM) Higgs boson with 5 fb$^{-1}$ data \cite{LHC}. Higgs bosons
have been excluded in a significant mass range and, ignoring the unlikely
possibility of a very heavy particle,  only the very narrow window $m_h \approx
115$--130 GeV is now left over. There is even a slight excess of events in the data which
could correspond to a SM like Higgs boson with  a mass of $125 \pm 1 $ GeV.
Although the statistics are not sufficient for the experiments to claim 
discovery, one is tempted to take this piece of evidence seriously and  analyze
its consequences.

In this Letter, we study the implications of these LHC  results for
Higgs-portal models of dark matter (DM). The Higgs sector of the SM enjoys a
special status since it allows for a direct  coupling  to the hidden sector that
is renormalizable. Hence,  determination of the properties of the Higgs boson
would allow us to gain information about the hidden world. The latter is
particularly important in the context of dark matter since hidden sector
particles can be stable and couple very weakly to the SM sector, thereby
offering a viable dark matter candidate \cite{Silveira:1985rk}. In principle,
the Higgs boson could  decay into light DM particles which escape detection
\cite{Shrock:1982kd}.  However, given  the fact that the ATLAS and CMS signal is
close to what one expects for a Standard Model--like Higgs particle,  there is
little room for invisible decays. In what follows, we will assume that 10\% is
the upper bound on the invisible Higgs decay branching ratio, although  values
up to 20\% will not significantly change our conclusions. 

We  adopt a model independent approach and study generic scenarios in which  the
Higgs-portal DM is a scalar, a vector or a Majorana fermion. We first discuss
the available  constraints on the thermal DM from WMAP and current direct
detection experiments, and show that the fermionic DM case is excluded while in
the scalar and vector cases, one needs  DM particles that are heavier than about
60 GeV. We then  derive the direct DM detection rates to be probed by the
XENON100--upgrade and XENON1T experiments. Finally, we  discuss the possibility
of observing directly or indirectly these DM particles in collider experiments
and, in particular, we determine the rate for the pair production of scalar 
particles at the LHC and a high-energy $e^+e^-$ collider. 

\section*{The models}

Following the model independent approach of Ref.~\cite{Kanemura:2010sh}, we 
consider the three possibilities that dark matter  consists of real scalars $S$,
vectors $V$ or Majorana fermions $\chi$   which interact with  the SM fields
only through the Higgs-portal. The stability of the DM particle is ensured by a $Z_2$
parity, whose origin is  model--dependent. For example, in the  vector case it
stems  from a natural parity symmetry of abelian gauge sectors with minimal
field content \cite{Lebedev:2011iq}.   The relevant  terms in the 
Lagrangians  are 
\begin{eqnarray} \!&&\Delta {\cal L}_S = -{1\over 2} m_S^2 S^2 - {1\over 4}
\lambda_S S^4 -  {1\over 4} \lambda_{hSS}  H^\dagger H  S^2 \;, \nonumber \\
\!&&\Delta {\cal L}_V = {1\over 2} m_V^2 V_\mu V^\mu\! +\! {1\over 4} \lambda_{V} 
(V_\mu V^\mu)^2\! +\! {1\over 4} \lambda_{hVV}  H^\dagger H V_\mu V^\mu ,
\nonumber \\ \! &&\Delta {\cal L}_f = - {1\over 2} m_f \bar \chi \chi -  {1\over 4}
{\lambda_{hff}\over \Lambda} H^\dagger H \bar \chi \chi \;.   \end{eqnarray}
Although in the fermionic  case above the Higgs--DM coupling is not
renormalizable,  we still include it for completeness. The self--interaction
terms $S^4$ in the scalar case and the $(V_\mu V^\mu)^2$ term in the vector 
case are not essential for our discussion and we will ignore them. After
electroweak symmetry breaking, the neutral component of the doublet field $H$ is
shifted to  $H^0 \rightarrow v + h/ \sqrt{2}$ with $v=174$ GeV and the physical
masses of the DM particles will be given by
\begin{eqnarray} && M_S^2 = m_S^2 + {1\over 2} \lambda_{hSS} v^2 \;, \nonumber \\
&&  M_V^2 = m_V^2 + {1\over 2} \lambda_{hVV} v^2 \;, \nonumber \\ && M_f = m_f +
{1\over 2} {\lambda_{hff}\over \Lambda} v^2 \;. \end{eqnarray}

In what follows, we summarize the most important formulas relevant to our study.
Related ideas and analyses can be found in  \cite{all1,Peter,Andreas:2010dz,
all2} and more  recent studies of Higgs-portal scenarios have appeared in
\cite{all3,Chu:2011be}.

The relic abundance of the DM particles is obtained through  the $s$--channel 
annihilation via the exchange of the Higgs boson. For instance, the  
annihilation cross section into light  fermions of mass $m_{\rm ferm}$  is given by 
\begin{eqnarray}
&& \langle  \sigma_{\rm ferm}^S v_r   \rangle = { \lambda_{hSS}^2 m_{\rm ferm}^2 \over 16 \pi }~
{1  \over  (4 M_S^2 - m_h^2)^2  }  \;, \nonumber\\
&& \langle  \sigma_{\rm ferm}^V v_r   \rangle = { \lambda_{hVV}^2 m_{\rm ferm}^2 \over 48 \pi }~
{1  \over  (4 M_V^2 - m_h^2)^2  }  \;, \nonumber\\
&& \langle  \sigma_{\rm ferm}^f v_r   \rangle = { \lambda_{hff}^2 m_{\rm ferm}^2 \over 32 \pi }~
{M_f^2 \over \Lambda^2}~ {v_r^2  \over  (4 M_f^2 - m_h^2)^2  } \;,
\end{eqnarray}
where $v_r$ is the DM relative velocity. (The cross section for Majorana fermion
annihilation was computed in \cite{McDonald:2008up} in a similar framework.) 
We
should note that in our numerical analysis,  we take into account the full set 
of relevant diagrams and channels, and we have adapted the program micrOMEGAs
\cite{Micromegas} to calculate the relic DM density.

The properties of the dark matter particles can be studied in direct detection
experiments.  The DM interacts elastically with nuclei through the Higgs boson
exchange. The  resulting nuclear recoil is then interpreted in terms of the DM
mass and  DM--nucleon cross section.  The spin--independent DM--nucleon
interaction can be expressed as \cite{Kanemura:2010sh}
\begin{eqnarray}
&& \sigma^{SI}_{S-N} = \frac{\lambda_{hSS}^2}{16 \pi m_h^4} \frac{m_N^4  f_N^2}{ (M_S + m_N)^2} \;,  \nonumber\\
&& \sigma^{SI}_{V-N} = \frac{\lambda_{hVV}^2}{16 \pi m_h^4} \frac{m_N^4  f_N^2}{ (M_V + m_N)^2} \;, \nonumber \\
&& \sigma^{SI}_{f-N} = \frac{\lambda_{hff}^2}{4 \pi \Lambda^2 m_h^4} \frac{m_N^4 M_f^2  f_N^2}{ (M_f + m_N)^2} \;,
\end{eqnarray}
where $m_N$ is the nucleon mass and $f_N$ parameterizes the Higgs--nucleon
coupling. The latter subsumes  contributions of the light quarks $(f_L)$ and
heavy quarks $(f_H)$, $f_N=\sum f_L + 3 \times \frac{2}{27} f_H$. There exist 
different estimations of this factor and in what follows we will use the lattice
result $f_N=0.326$ \cite{Young:2009zb} as well as the MILC results
\cite{Toussaint:2009pz} which provide  the minimal value $f_N=0.260$ and the
maximal value $f_N=0.629$. We note that the most recent lattice evaluation of
the strangeness content of the nucleon \cite{Bali:2011rs} favors $f_N$ values 
closer to the lower end of the above range. 
In our numerical analysis, we have taken into account these lattice results,
which appear more reliable than those
extracted from the pion--nucleon cross section.

If the DM particles are light enough, $M_{\rm DM} \leq \frac12 m_h$, they will
appear as  invisible decay products of the Higgs boson. For the various cases, 
the Higgs partial decay  widths into invisible DM particles are given by
\begin{eqnarray}
 &&\Gamma^{{\rm inv}}_{h\rightarrow SS} = \frac{\lambda_{hSS}^2 v^2 \beta_S}{64 \pi  m_h}  \;, \nonumber\\
&& \Gamma^{\rm inv}_{h \rightarrow V V} = \frac{\lambda^2_{hVV} v^2 m_h^3
\beta_V }{256 \pi   M_{V}^4}
\left( 1-4 \frac{M_V^2}{m_h^2}+12\frac{M_V^4}{m_h^4}
\right), \nonumber\\
&&  \Gamma^{\rm inv}_{h \rightarrow \chi \chi} = {\lambda_{hff}^2 v^2 m_h 
\beta_f^{3}\over 32 \pi 
\Lambda^2 }  \;, 
\end{eqnarray}
where $\beta_X=\sqrt{1-4M_X^2/m_h^2}$. We have adapted the program HDECAY 
\cite{Hdecay} which calculates all Higgs decay widths and branching ratios
to include invisible decays. 

\section*{Astrophysical  consequences}

The first aim of our study is to derive constraints on the various DM particles
from the  WMAP satellite \cite{WMAP} and from the current direct detection 
experiment XENON100 \cite{Aprile:2011ts}, and to make predictions  for future
upgrades of the latter experiment, assuming that  the  Higgs boson  has a mass
$m_h =125$ GeV and is approximately SM--like such  that its invisible decay 
branching ratio is smaller than 10\%; we have checked that increasing this
fraction  to 20\%  does not change our results significantly.

\begin{figure}[!h]
    \begin{center}
    \hspace{-1.cm}
   \includegraphics[width=2.4in]{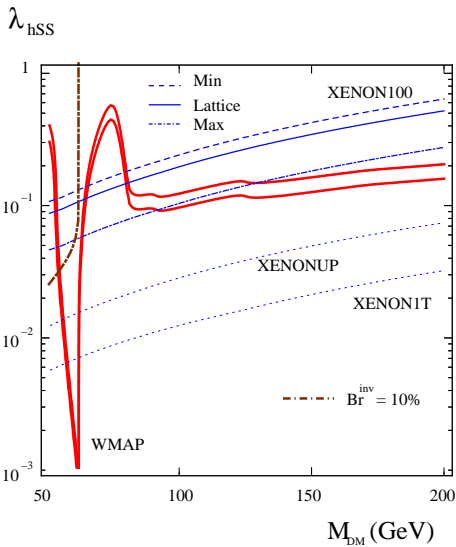}
\vspace*{-2mm}
          \caption{{\footnotesize
Scalar Higgs-portal 
parameter space allowed by WMAP (between the solid red curves),  
XENON100  and ${\rm BR}^{\rm inv}\!=\! 10\%$ 
  for $m_h\!=\! 125$ GeV. Shown also are the prospects for XENON upgrades. 
}}
\label{Fig:SCANSCAL}
\end{center}
\vspace*{-9mm}
\end{figure}

In Fig.~\ref{Fig:SCANSCAL}, we delineate the viable  parameter space for the
Higgs-portal scalar DM particle.   The area between the two  solid (red) curves
satisfies the WMAP constraint, with the dip corresponding to resonant DM
annihilation mediated by the Higgs exchange. We display three versions of the
XENON100 direct DM detection bound corresponding to the three  values of $f_N$
discussed above.  The dash--dotted  (brown) curve around the Higgs pole region 
represents ${\rm BR}^{\rm inv}=10\%$ such that  the area to the left of this
line is excluded by our constraint ${\rm BR}^{\rm inv} < 10\%$.  The prospects
for the upgrade of XENON100 (with a projected sensitivity corresponding to  
60,000 kg-d, 5-30 keV and 45\% efficiency)   and XENON1T are shown   by the
dotted lines.

We find that light dark matter,  $M_{\rm DM} \lsim 60 $ GeV, violates the bound
on the invisible Higgs decay branching ratio and thus is excluded. This applies
in particular to the case of  scalar DM with a mass of 5--10 GeV considered, 
for instance, in  Ref.~\cite{Andreas:2010dz}. On the other hand, heavier dark
matter, particularly for $M_{\rm DM} \gsim 80 $ GeV,  is allowed  by both  
${\rm BR}^{\rm inv}$ and XENON100.  We note that almost the entire available
parameter space will be probed by the XENON100 upgrade. The exception is a small
resonant region around 62 GeV, where the Higgs--DM coupling is extremely  small.

\begin{figure}[!h]
    \begin{center}
    \hspace{-1.cm}
   \includegraphics[width=2.4in]{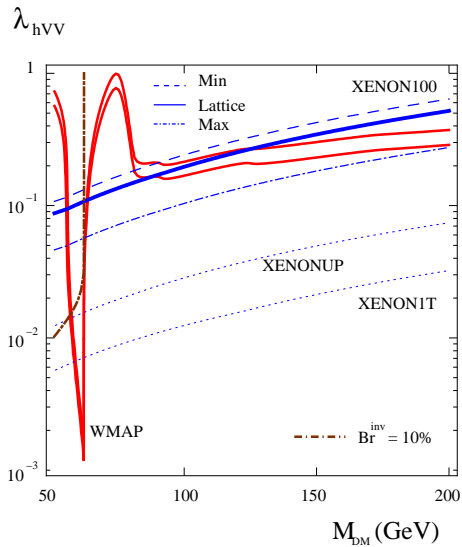}
\vspace*{-2mm}
 \caption{{\footnotesize
Same as Fig.~1 for vector DM particles.
}}
\label{Fig:SCANVEC}
\end{center}
\vspace*{-4mm}
\end{figure}

In the case of  vector Higgs-portal DM, the results are  shown in
Fig.~\ref{Fig:SCANVEC} and are  quite similar to the scalar case. WMAP requires
the Higgs--DM coupling to be almost twice as large as  that in the scalar case.
This is because only opposite polarization states can annihilate through the
Higgs channel, which reduces the annihilation cross section by a factor of 3.
The resulting direct detection rates are therefore somewhat higher in the vector
case. Note that  for DM masses below $m_h/2$, only  very small values
$\lambda_{hVV}\! <\! {\cal O}(10^{-2})$ are allowed if ${\rm BR}^{\rm inv}\! <\!
10\%$.

\begin{figure}[!]
    \begin{center}
    \hspace{-1.cm}
   \includegraphics[width=2.4in]{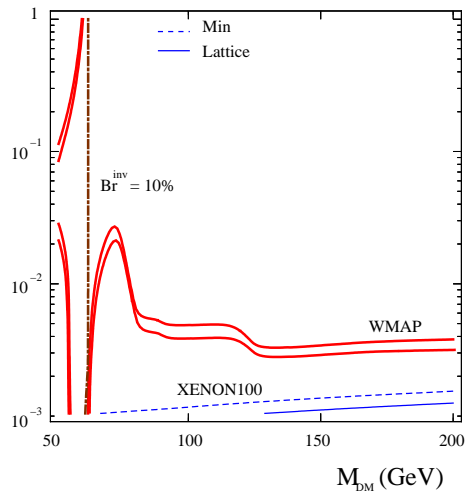}
 \caption{{\footnotesize
Same as in Fig.1 for fermion DM;  $\lambda_{hff}/\Lambda$ is 
in ${\rm GeV}^{-1}$.  
}}
\label{Fig:SCANFERM}
\end{center}
\vspace*{-7mm}
\end{figure}

Similarly, the fermion Higgs-portal results are shown  in
Fig.~\ref{Fig:SCANFERM}. We find no parameter regions satisfying the
constraints, most notably the XENON100 bound, and this scenario is thus 
ruled out for $\lambda_{hff}/\Lambda \gsim 10^{-3}$.

\begin{figure}[!h]
    \begin{center}
    \hspace{-.4cm}
   \includegraphics[width=3.5in]{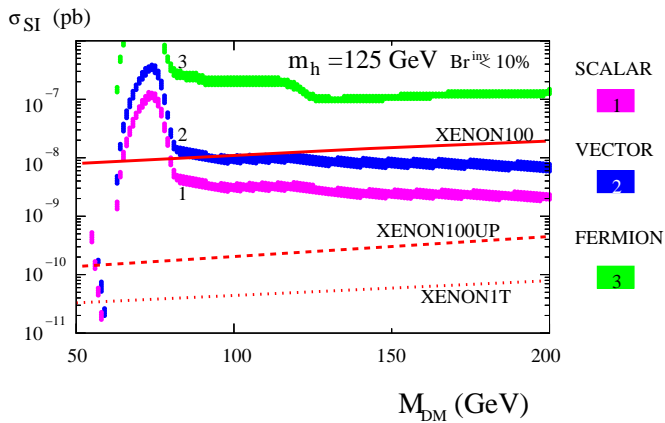}
          \caption{{\footnotesize
Spin independent DM--nucleon cross section 
versus DM mass.
The upper band (3)  corresponds to fermion DM, the middle one (2)
to vector DM and the lower one (1)   to scalar DM. The solid, 
dashed and dotted lines represent XENON100, XENON100 upgrade
and XENON1T sensitivities, respectively.
 }}
\label{Fig:SigmaSIbis}
\end{center}
\vspace*{-5mm}
\end{figure}

This can also be seen from  Fig.~\ref{Fig:SigmaSIbis}, which displays
predictions for the spin--independent  DM--nucleon cross section  $\sigma_{\rm
SI}$ (based on the lattice $f_N$)    subject to the WMAP and  ${\rm BR}^{\rm
inv} < 10\%$ bounds. The upper band  corresponds to the fermion Higgs-portal DM
and is excluded by XENON100. On the other hand, scalar and vector DM are both
allowed for a wide range of masses. Apart from a very small region around
$\frac12 m_h$, this parameter space will be probed by  XENON100--upgrade and XENON1T. The
typical value for the scalar  $\sigma_{\rm SI }$ is a few times $10^{-9}$ pb,
whereas $\sigma_{\rm SI }$ for vectors is larger by a factor  of 3 which
accounts for the number of  degrees of freedom.

\section*{Dark matter production at colliders}
\vspace*{-1mm}

The next issue to discuss is how to observe  directly the Higgs-portal DM
particles at high energy colliders. There are essentially two ways,  depending
on the Higgs versus DM particle masses. If the DM particles are light enough
for  the invisible Higgs decay to occur,  $M_{\rm DM} \lsim \frac12 m_h$, the
Higgs cross sections times the branching ratios for the visible decays will be
altered, providing indirect evidence for the invisible decay channel. In the
case of the LHC, a detailed analysis of this issue has been performed in
Ref.~\cite{Peter} for instance and we have little to add to it. Nevertheless, if
the invisible Higgs branching ratio is smaller than $\approx 10\%$, its
observation would be extremely difficult   in view of the large QCD
uncertainties that affect the Higgs production cross sections, in particular in
the main  production channel, the gluon fusion mechanism $gg \to h$ \cite{JB}.
In fact, the chances of observing indirectly the invisible Higgs decays are much
better at a future $e^+e^-$ collider. Indeed,  it has been shown  that, at
$\sqrt s \approx 500$ GeV collider  with 100 fb$^{-1}$ data,  the Higgs
production cross sections times the visible decay branching  fractions can be
determined at the percent level \cite{TESLA,Review2}. 

The DM particles could be observed directly by studying associated Higgs 
production with a vector boson  and Higgs production in  vector boson fusion 
with the Higgs particle  decaying invisibly. At the LHC, parton level analyses
have shown that, although extremely difficult, this channel can be probed  at
the 14 TeV upgrade with a sufficiently large amount of data \cite{LHC-inv} if
the fraction of invisible decays is significant. A  more sophisticated  ATLAS
analysis has shown that only for branching ratios above 30\% that a signal can
be observed at $\sqrt s= 14$ TeV and 10 fb$^{-1}$ data in the mass range
$m_h=100$--250 GeV \cite{LHC-ATLAS}.  Again, at a 500 GeV  $e^+e^-$ collider,
invisible decays  at the level of a few percent can be observed in the process
$e^+e^- \to hZ$ by simply analyzing the recoil of the leptonically decaying $Z$
boson \cite{TESLA,Review2}.

If the DM particles are heavy, $M_{\rm DM} \gsim \frac12 m_h$, the situation  
becomes much more difficult and the only possibility to observe them would be
via  their  pair production  in the continuum through the $s$--channel exchange
of the Higgs boson. At the LHC, taking the example of the scalar DM  particle
$S$, three main processes can be used:  $a)$ double production with
Higgs--strahlung from either a $W$ or a  $Z$ boson, $q\bar{q} \to  V^* \to VSS$
with $V=W$ or $Z$,  $b)$ the $WW/ZZ$ fusion processes which lead to two jets and
missing energy $qq \to V^* V^* qq  \to SSqq$ and $c)$ the gluon--gluon fusion
mechanism which is mainly mediated by loops of the heavy top quark  that couples
strongly to the Higgs boson, $gg \to h^* \to SS$. 

The third process, $gg\to SS$, leads to only invisible particles in the final
state, unless some additional jets from higher order contributions are present
and reduce the cross section \cite{Hgg} and we will ignore it here. For the two
first processes, following Ref.~\cite{HHH} in which double Higgs production in
the SM and its minimal supersymmetric extension  has been analyzed, we have
calculated the production  cross sections. The exact matrix elements  have been
used in the $q\bar q \to ZSS,WSS$ processes while in vector boson fusion, we
have used the longitudinal vector boson approximations and specialized to the 
$W_L W_L+Z_LZ_L \to SS$  case which is expected to provide larger rates at the
highest energy  available at the LHC i.e. $\sqrt s\!=\!14$ TeV (the result
obtained in this way is expected to approximate the exact result  within about a
factor of two for low scalar masses and very high energies);  the analytical
expressions are given in the Appendix.\s

As can be seen from Fig.~\ref{Fig:LHC} where the cross sections are shown as a
function of $M_{\rm DM}$ for $\lambda_{hSS}= 1$, the rates at $\sqrt s=14$ TeV
are at the level of 10 fb in the $WW+ZZ\to SS$ process for $M_h \lsim 120$ GeV
and one order of magnitude smaller for associated production with $W$ and $Z$
bosons. Thus, for both processes, even before selection cuts are applied to
suppress the backgrounds, the rates are small for DM masses of order  100 GeV
and will require extremely high luminosities to be observed. 


\begin{figure}
    \begin{center}
   \includegraphics[width=3in]{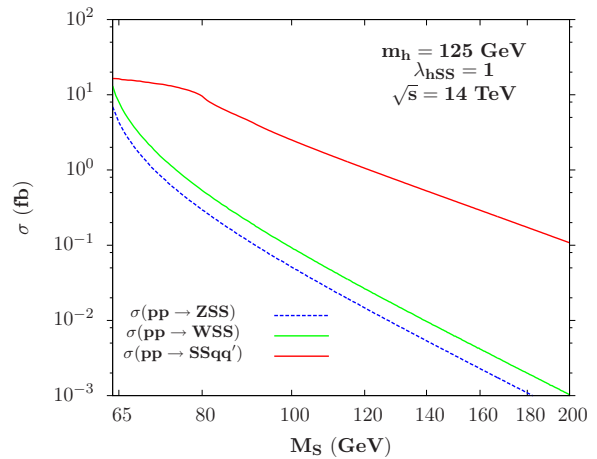}
 \caption{{\footnotesize
Scalar DM pair production cross sections at the LHC with $\sqrt s=14$ TeV
as a function of their mass for $\lambda_{hSS}=1$ in the processes $pp \to
ZSS,WSS$ and $pp \to W^*W^*\!+\!Z^*Z^*\to SSqq$.
}}
\end{center}
\end{figure}

\begin{figure}[!]
    \begin{center}
    \vspace{-.2cm}
   \includegraphics[width=3in]{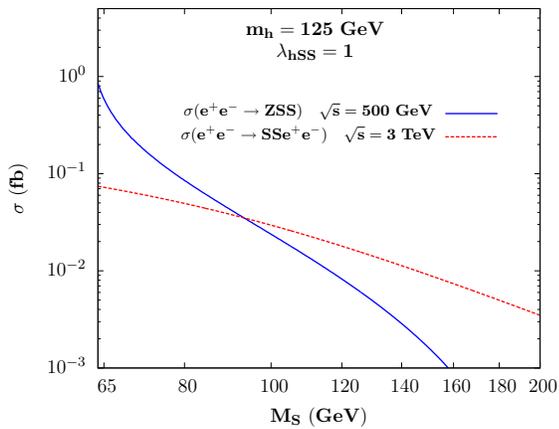}
\vspace*{-5.3cm}
 \caption{{\footnotesize
Scalar DM pair production cross sections at $e^+ e^-$ colliders  as a function
of the DM mass for $\lambda_{hSS}\!=\!1$ in the processes $e^+ e^- \to ZSS$ at
$\sqrt s=500$ GeV and  $ZZ \to SS$ at $\sqrt s=3$ TeV.
}}
\label{Fig:ILC}
\end{center}
\vspace*{-6mm}
\end{figure}

Again, the chances of observing DM pair production in the continuum  might be
higher in the cleaner environment of $e^+ e^-$ collisions. The two most
important production processes in this context, taking again the example of a
scalar DM particle, are   $e^+ e^- \to ZSS$ that dominates at relatively low
energies and $e^+ e^- \! \! \to Z^* Z^* e^+ e^- \! \to \! e^+ e^-SS$  which 
becomes important at high energies.  The rate for $WW$ fusion is one order of
magnitude larger but it leads to a fully invisible signal, $e^+ e^-\!  \to \!
W^* W^* \nu \bar \nu \! \to \! \nu \bar \nu SS$. Following again Ref.~\cite{HHH}, we
have  evaluated  the cross sections  for $e^+ e^- \to ZSS$ at $\sqrt s=500$ GeV
(the  energy range relevant for the ILC) and for $Z_L Z_L \to SS$ at  $\sqrt s=
3$ TeV (relevant for the CERN CLIC) and the results are shown in
Fig.\ref{Fig:ILC} as a function of the mass $M_S$  for $\lambda_{hSS}=1$. One
observes that the maximal rate that one can obtain is about 10 fb near the Higgs
pole in $ZSS$ production and which drops quickly with increasing $M_S$. The process
$ZZ\to SS$ becomes  dominant for $M_S \gsim 100$ GeV, but the rates are
extremely low, below $\approx 0.1$ fb.  

The situation should be similar in the case of vector and  fermion DM and we
refrain from discussing it  here. 

\section*{Conclusion}

We have analyzed the implications of the recent LHC Higgs results for  generic
Higgs-portal models of  scalar, vector and fermionic dark matter particles.
Requiring the branching ratio for invisible Higgs decay to be less than 10\%,
we  find that the DM--nucleon cross section for electroweak--size DM masses is
predicted to be in the range $10^{-9}-10^{-8}$ pb  in almost all of the 
parameter space. Thus, the entire class of Higgs-portal DM models   will be
probed by the XENON100--upgrade and XENON1T direct detection experiments, which
will also be able to discriminate between the vector and scalar cases.  The
fermion DM is essentially ruled out by the current data, most notably by
XENON100. Furthermore, we  find that light Higgs-portal    DM  $M_{\rm DM} \lsim
60$ GeV is  excluded independently of its nature since it predicts  a large
invisible Higgs decay branching ratio, which should be incompatible with the 
production of an SM--like Higgs boson at the LHC.   Finally, it will be difficult
to observe the DM effects by studying Higgs physics at the LHC. Such studies  
can be best performed in Higgs decays at the planned $e^+e^-$ colliders. 
However, the  DM particles have pair production cross sections that are too low
to be observed at the LHC and eventually also at future  $e^+e^-$ colliders
unless very high luminosities are made available. \smallskip

\noindent {\bf Acknowledgements. }  The authors would like  to thank E. Bragina,
J.B. de Vivie and M. Kado for discussions as well  as  S. Pukhov for his help in
solving technical problems related to micrOMEGAs.  This  work was supported by
the French ANR TAPDMS {\bf ANR-09-JCJC-0146}  and the Spanish MICINNÕs
Consolider-Ingenio 2010 Programme  under grant  Multi- Dark {\bf CSD2009-00064}.

\subsection*{Appendix}
\setcounter{equation}{0}
\renewcommand{\theequation}{A.\arabic{equation}}

The differential cross section for the pair production of two scalar
particles in association with a $Z$ boson,  $e^+e^-  \to ZSS$, after the
angular dependence is integrated out, can be cast into the form
($v=174$ GeV):
\beq 
\frac{{\rm d} \sigma (e^+ e^- \to ZSS)}{{\rm d} x_1 {\rm d} x_2} = 
\frac{G_F^3 M_Z^2 v^4}{384 \sqrt{2} \pi^3 s}
\frac{(\hat a_e^2 + \hat v_e^2)}{(1- \mu_Z)^2}\, \lambda_{hSS}^2\, {\cal Z} \; ,
\eeq 
where the electron--$Z$ couplings are defined as $\hat a_e=-1$ and $\hat v_e=
-1+4 \sin^2\theta_W$,  $x_{1,2} =2 E_{1,2}/\sqrt{s}$ are the scaled energies of the two
scalar  particles, $x_3 = 2 - x_1 -x_2$ is the scaled energy of the $Z$ boson; 
the scaled masses are denoted by $\mu_i =
M_i^2/s$. In terms of these variables, the coefficient ${\cal Z}$ may be written
as
\beq 
{\cal Z} =\frac{1}{4} \frac{ \mu_Z (x_3^2+8\mu_Z) }
{(1-x_3+\mu_Z-\mu_h)^2} \; .
\eeq
The differential cross section has to be integrated over the allowed range of 
the $x_1, x_2$ variables; the boundary condition is 
\begin{equation}
\left| \frac{2(1-x_1-x_2+2\mu_S-\mu_Z) + x_1x_2}
{\sqrt{x_1^2-4\mu_S} \sqrt{x_2^2-4\mu_S}} \right| \leq 1  \;.
\label{eq:dalitzbound}
\end{equation}
For the cross section at hadron colliders, i.e. for the process $q\bar q \to
ZSS$ one has to divide the amplitude squared given above by a factor 3 to take
into account color sum/averaging, replace $e$ by $q$ (with $a_q= 2I_q^3, v_q =
2I_q^3-4e_q \sin^2\theta_W$ with $I_q^3$ and $e_q$ for isospin and electric 
charge) and  the c.m. energy $s$ by the partonic one $\hat s$; one has  then to
fold the obtained partonic cross section with the quark/antiquark luminosities.
The extension to the $q \bar q \to WSS$ case  (with $a_q=v_q= \sqrt 2$) is
straightforward. 

For the vector boson fusion processes,  one   calculates the cross sections for 
the $2 \to 2$ processes $V_L V_L \to SS$  in the equivalent longitudinal vector
boson approximation and then fold with the $V_L$ spectra to obtain the cross
section the  entire processes $e^+ e^- \to SS \ell \ell $ and $qq\to qq SS$;
see Ref.~\cite{HHH} for details. 
Taking into account only the dominant longitudinal vector boson contribution,  
denoting by $\beta_{V,S}$ the $V,S$ velocities in the c.m.\ frame, 
$\hat{s}^{1/2}$ the invariant energy of the $VV$ pair, the corresponding  
cross section of the subprocess $V_L V_L \to SS$  reads
\beq
\hat{\sigma}_{V_LV_L} = \frac{G_F^2 M_V^4 v^4}{4\pi \hat{s}}
\lambda_{hSS}^2 \, \frac{\beta_S}{\beta_W}
\left[ \frac{1+\beta_W^2}{1-\beta_W^2}  \frac{1}
{(\hat{s} -M_h^2)} \right]^2 \; .
\eeq
The result obtained after folding with the vector boson spectra is expected to
approximate the exact result  within about a factor of two for low scalar masses
and very high energies.


\end{document}